\documentclass[fleqn,usenatbib]{mnras}

\usepackage{newtxtext,newtxmath}
\usepackage[T1]{fontenc}
\usepackage{lineno}

\usepackage{enumitem}
\newcounter{descriptcount}
\newlist{enumdescript}{description}{1}
\setlist[enumdescript,1]{%
  before={\setcounter{descriptcount}{0}%
          \renewcommand*\thedescriptcount{\arabic{descriptcount}}},
        font={\bfseries\stepcounter{descriptcount}Case \thedescriptcount,~}
}

\makeatletter
\newcommand*{\mint}[1]{%
  \mint@l{#1}{}%
}
\newcommand*{\mint@l}[2]{%
  \@ifnextchar\limits{%
    \mint@l{#1}%
  }{%
    \@ifnextchar\nolimits{%
      \mint@l{#1}%
    }{%
      \@ifnextchar\displaylimits{%
        \mint@l{#1}%
      }{%
        \mint@s{#2}{#1}%
      }%
    }%
  }%
}
\newcommand*{\mint@s}[2]{%
  \@ifnextchar_{%
    \mint@sub{#1}{#2}%
  }{%
    \@ifnextchar^{%
      \mint@sup{#1}{#2}%
    }{%
      \mint@{#1}{#2}{}{}%
    }%
  }%
}
\def\mint@sub#1#2_#3{%
  \@ifnextchar^{%
    \mint@sub@sup{#1}{#2}{#3}%
  }{%
    \mint@{#1}{#2}{#3}{}%
  }%
}
\def\mint@sup#1#2^#3{%
  \@ifnextchar_{%
    \mint@sup@sub{#1}{#2}{#3}%
  }{%
    \mint@{#1}{#2}{}{#3}%
  }%
}
\def\mint@sub@sup#1#2#3^#4{%
  \mint@{#1}{#2}{#3}{#4}%
}
\def\mint@sup@sub#1#2#3_#4{%
  \mint@{#1}{#2}{#4}{#3}%
}
\newcommand*{\mint@}[4]{%
  \mathop{}%
  \mkern-\thinmuskip
  \mathchoice{%
    \mint@@{#1}{#2}{#3}{#4}%
        \displaystyle\textstyle\scriptstyle
  }{%
    \mint@@{#1}{#2}{#3}{#4}%
        \textstyle\scriptstyle\scriptstyle
  }{%
    \mint@@{#1}{#2}{#3}{#4}%
        \scriptstyle\scriptscriptstyle\scriptscriptstyle
  }{%
    \mint@@{#1}{#2}{#3}{#4}%
        \scriptscriptstyle\scriptscriptstyle\scriptscriptstyle
  }%
  \mkern-\thinmuskip
  \int#1%
  \ifx\\#3\\\else_{#3}\fi
  \ifx\\#4\\\else^{#4}\fi  
}
\newcommand*{\mint@@}[7]{%
  \begingroup
    \sbox0{$#5\int\m@th$}%
    \sbox2{$#5\int_{}\m@th$}%
    \dimen2=\wd0 %
    \let\mint@limits=#1\relax
    \ifx\mint@limits\relax
      \sbox4{$#5\int_{\kern1sp}^{\kern1sp}\m@th$}%
      \ifdim\wd4>\wd2 %
        \let\mint@limits=\nolimits
      \else
        \let\mint@limits=\limits
      \fi
    \fi
    \ifx\mint@limits\displaylimits
      \ifx#5\displaystyle
        \let\mint@limits=\limits
      \fi
    \fi
    \ifx\mint@limits\limits
      \sbox0{$#7#3\m@th$}%
      \sbox2{$#7#4\m@th$}%
      \ifdim\wd0>\dimen2 %
        \dimen2=\wd0 %
      \fi
      \ifdim\wd2>\dimen2 %
        \dimen2=\wd2 %
      \fi
    \fi
    \rlap{%
      $#5%
        \vcenter{%
          \hbox to\dimen2{%
            \hss
            $#6{#2}\m@th$%
            \hss
          }%
        }%
      $%
    }%
  \endgroup
}

\DeclareRobustCommand{\VAN}[3]{#2}
\let\VANthebibliography\thebibliography
\def\thebibliography{\DeclareRobustCommand{\VAN}[3]{##3}\VANthebibliography}

\usepackage{graphicx}	% Including figure files
\usepackage{amsmath}	% Advanced maths commands
\title[The Wave Energy Density and Growth Rate]{The Wave Energy Density and Growth Rate for the Resonant Instability in Relativistic Plasmas}

\author[S.-Y. Jeong et al.]{
Seong-Yeop Jeong,$^{1}$\thanks{E-mail: kotr7724@gmail.com}
and Clare Watt,$^{1}$
\\
$^{1}$Northumbria University, Newcastle, NE1 8ST, UK\\
}

\date{Accepted XXX. Received YYY; in original form ZZZ}

\pubyear{2022}

\begin{document}
\label{firstpage}
\pagerange{\pageref{firstpage}--\pageref{lastpage}}
\maketitle

\begin{abstract}
The wave instability acts in astrophysical plasmas to redistribute energy and momentum in the absence of frequent collisions. There are many different types of waves, and it is important to quantify the wave energy density and growth rate for understanding what type of wave instabilities are possible in different plasma regimes. There are many situations throughout the universe where plasmas contain a significant fraction of relativistic particles. Theoretical estimates for the wave energy density and growth rate are constrained to either field-aligned propagation angles, or non-relativistic considerations. Based on linear theory, we derive the analytic expressions for the energy density and growth rate of an arbitrary resonant wave with an arbitrary propagation angle in relativistic plasmas. For this derivation, we calculate the Hermitian and anti-Hermitian parts of the relativistic-plasma dielectric tensor. We demonstrate that our analytic expression for the wave energy density presents an explicit energy increase of resonant waves in the wavenumber range where the analytic expression for the growth rate is positive (i.e., where a wave instability is driven). For this demonstration, we numerically analyse the loss-cone driven instability, as a specific example, in which the whistler-mode waves scatter relativistic electrons into the loss cone in the radiation belt. Our analytic results further develop the basis for linear theory to better understand the wave instability, and have the potential to combine with quasi-linear theory, which allows to study the time evolution of not only the particle momentum distribution function but also resonant wave properties through an instability.
\end{abstract}


%%%%%%%%%%%%%%%%%%%%%%%%%%%%%%%%%%%%%%%%%%%%%%%%%%

%%%%%%%%%%%%%%%%% BODY OF PAPER %%%%%%%%%%%%%%%%%%

\section{Introduction}\label{intro}

The wave instability is a physical mechanism, in which the unstable wave resonates with charged particles at a specific velocity, and grows with time by absorbing the particle's kinetic energy. It plays a crucial role for the stabilisation of charged particles in many space and astrophysical plasmas. For example, the wave instability contributes to the isotropisation of the loss-cone momentum distribution function (MDF) of relativistic electrons in the Earth's radiation belt \citep{Ukhorskiy2014}, the heat-flux regulation in the intracluster medium of galaxy clusters \citep{Roberg_Clark_2016}, and the scattering of the surface radiation in the relativistic electron-positron plasma of neutron-star magnetospheres \citep{10.1111/j.1365-2966.2006.10140.x,article}. Relativistic collisionless shocks lead to plasma environments with many different types of wave instabilities \citep{Nakar_2011,Marcowith_2016}. In addition, wave instabilities are likely responsible for the isotropisation of solar wind plasmas \citep{10.3389/fphy.2021.624748}. Therefore, it is of great importance to study the wave instability in relativistic plasmas in order to advance our understanding of the wave-particle interaction throughout the universe. 

Relativistic plasma environments exist in easily-accessible near-Earth regions, for example. We can use this specific environment to test theoretical ideas about kinetic wave-particle interactions, because the space-based instrumentation allows \textit{in-situ} measurements for plasmas and waves. There are many types of kinetic wave instabilities in the Earth's magnetsophere and magnetosheath, driven unstable by the presence of free energy in the electron or ion MDFs. Whistler-mode waves are often unstable in Earth's magnetosphere due to the presence of temperature anisotropy in electrons \citep{https://doi.org/10.1029/2009JA014845} or the atmospheric loss-cone \citep{https://doi.org/10.1029/JZ071i001p00001}. They interact with electrons over a wide range of energies and are important for the pitch-angle scattering and energisation. Note that although we know a lot about whistler-mode waves from studies of the Earth's magnetosphere, they have also been identified in the solar wind \citep{Tong_2019} where they are important for the pitch-angle scattering of the electron strahl \citep{Vocks_2003,Vasko_2019,Verscharen_2019,Jeong_2020,Jeong_2022}.

Even though whistler-mode waves, propagating parallel with respect to the background magnetic field in the magnetosphere, most efficiently experience the growth rate according to linear theory, whistler-mode waves propagate significant distances along the field line as they grow. In the inhomogeneous magnetic field of the inner magnetosphere, the spectra of whistler-mode waves therefore have a mix of the parallel and oblique propagation \citep{https://doi.org/10.1002/jgra.50231,https://doi.org/10.1029/2012JA018343}. The obliquely propagating whistler-mode waves are observed throughout Earth’s inner magnetosphere \citep{https://doi.org/10.1002/jgra.50312,https://doi.org/10.1002/jgra.50176,articleArtemyev}. Whistler-mode waves change the propagation angle as they refract through the magnetosphere \citep{2008JGRA..113.9210L}, leading to changes in local growth rates, and thus their local energy densities. 

These relativistic plasma environments are pervasive throughout the universe \citep{dorman2020space}. Therefore, it is important to advance our ability for studying the instabilities through the waves not only parallel but also obliquely propagating with respect to the background magnetic field in relativistic plasmas. 

For studying the wave instability, the wave energy density and growth rate as a function of the wavenumber or frequency are important milestones. Based on the weak-growth rate approximation, the plasma dispersion relation determines the analytic expression for the growth rate of an arbitrary wave with an arbitrary propagation angle in non-relativistic plasmas \citep{kennel_wong_1967}. Although this analytic expression for the growth rate must be generalised by including the relativistic effect, it is not analytically completed \citep{1973ApJ...184..251B,1981A&A....98..161A,1986NCimD...8..318O}. Alternatively, based on the assumption that the cold plasma dominates over the relativistic and unstable plasma, the wave growth rate is calculated in relativistic plasmas \citep{doi:10.1063/1.872932,https://doi.org/10.1029/2009JA014428,doi:10.1063/1.5089749}. On the other hand, the general and analytic expression for the energy density of the resonant wave is still left unclear \citep{1992wapl.book.....S}. Only through quasi-linear theory, the time evolution for the energy density of the parallel-propagating resonant wave is estimated under the action of an instability \citep{doi:10.1063/1.1692186,2017RvMPP...1....4Y,Shaaban_2021}. To better understand the physics of the wave instability in relativistic plasmas, the analytic expression for the energy density of resonant waves as a clear form is essential.

The purpose of this paper is to derive clear expressions for the resonant wave energy density and growth rate based on the relativistic-plasma dielectric tensor, which are valid for all wave modes with all propagation angles in any gyrotropic MDF of relativistic plasmas. The fact is that the analytic expression for the wave energy density must present an energy density increase within a specific wavenumber range where the analytic expression for the growth rate presents positive values (i.e., where a wave instability is driven and the resonant wave grows). We demonstrate that the behaviours of these two expressions are correlated in such way, by numerically analysing the loss-cone driven instability, as a specific and readily observed example. Therefore, this paper provides analytical results, further developing the basis for linear theory. This paper also allows to combine linear and quasi-linear theories to study the time evolution of not only the particle MDF, accompanying non-Maxwellian features, but also resonant wave properties through a wave instability. This paper will play an important role in advancing our understanding of wave and plasma observations, and also kinetic simulation results from wave instabilities.

In Section~\ref{2}, we present the derivation of the analytic expressions for the energy density and growth rate of the resonant wave in relativistic plasmas. In Section~\ref{3}, to prove the correlation between these two expressions, we numerically investigate the loss-cone driven instability. In Section~\ref{4}, we discuss and conclude our results.

\section{Derivation} \label{2}

In this section, we explicitly derive the analytic expressions for the energy density and growth rate of an arbitrary resonant wave through an instability in relativistic plasmas. According to Poynting's theorem, the energy density and growth rate of the wave, resonating with charged particles, are given as \citep{1992wapl.book.....S}
\begin{equation}\label{waveenergy}
    W=\frac{1}{16\pi}\left[\mathbf{{B}}_k^{\ast} \cdot \mathbf{{B}}_k + \mathbf{{E}}_k^{\ast} \cdot \frac{\partial}{\partial \omega_k}\left(\omega_k \overleftrightarrow {\mathbf{K}}^h\right) \cdot \mathbf{{E}}_k \right],
\end{equation}
and
\begin{equation}\label{growthrate}
    \gamma_k=i\frac{\omega_k}{16\pi}\frac{\mathbf{{E}}_ k^{\ast}\cdot  \overleftrightarrow {\mathbf{K}}^a  \cdot  \mathbf{{E}}_ k }{W},
\end{equation}
where
\begin{equation}\label{hermitian}
    \overleftrightarrow {\mathbf{K}}^h=\frac{1}{2}\left(\overleftrightarrow {\mathbf{K}}+\overleftrightarrow {\mathbf{K}}^{\dag} \right),
\end{equation}
and
\begin{equation}\label{antihermitian}
    \overleftrightarrow {\mathbf{K}}^a=\frac{1}{2}\left(\overleftrightarrow {\mathbf{K}}-\overleftrightarrow {\mathbf{K}}^{\dag} \right).
\end{equation}
We denote the frequency as $\omega$, which is a complex function of the wavevector $\mathbf{k}$, and we define $\omega_{k}$ as its real part and $\gamma_k$ as its imaginary part ($\omega=\omega_{k}+i\gamma_k$). In our analysis, we take the local background magnetic field as $\mathbf{B}_0=\hat {\mathbf e}_z B_0$. The wavevector is denoted as $\mathbf{k}=(k_\perp,\phi_k,k_\parallel)$ where $k_x=k_\perp \cos\phi_k$, $k_y=k_\perp \sin\phi_k$ and $k_z=k_\parallel$. The first-order perturbed electromagnetic fields in plasmas are denoted as $\mathbf{E}_{k}(\omega, \mathbf k) =\hat {\mathbf e}_x E_k^x+\hat {\mathbf e}_y E_k^y+\hat {\mathbf e}_z E_k^z$ and $\mathbf{B}_{k}(\omega, \mathbf k)=\hat {\mathbf e}_x B_k^x+\hat {\mathbf e}_y B_k^y+\hat {\mathbf e}_z B_k^z$, which are spatially and temporally Fourier-transformed. We denote the dielectric tensor as
\begin{equation}\label{basicdielectrictensor}
\begin{split}
    \overleftrightarrow {\mathbf{K}}\equiv
\begin{bmatrix}
K_{xx}   & K_{xy}   & K_{xz}  \\
K_{yx}   & K_{yy}   & K_{yz}  \\
K_{zx}   & K_{zy}   & K_{zz}  \\
\end{bmatrix},
\end{split}
\end{equation}
and $\overleftrightarrow {\mathbf{K}}^h$ and $\overleftrightarrow {\mathbf{K}}^a$ are the Hermitian and anti-Hermitian parts of $\overleftrightarrow {\mathbf{K}}$. The superscript $*$ and $\dag$ indicate the conjugate and conjugate transpose, respectively. By definitions of the Hermitian and anti-Hermitian parts, Eqs.~(\ref{hermitian}) and (\ref{antihermitian}) prove that
%\begin{equation}\label{totalDtensor}
%    \overleftrightarrow {\mathbf{K}}=\overleftrightarrow {\mathbf{K}}^h+\overleftrightarrow {\mathbf{K}}^a,
%\end{equation}
$\overleftrightarrow {\mathbf{K}}^h=(\overleftrightarrow {\mathbf{K}}^h)^{\dag}$ and $\overleftrightarrow {\mathbf{K}}^a=-(\overleftrightarrow {\mathbf{K}}^a)^{\dag}$. Note that the dielectric tensor $\overleftrightarrow {\mathbf{K}}$ is based on the assumption that the plasma is homogeneous in space.

\subsection{Dielectric Tensor in Relativistic Plasmas}

We start our derivation from analysing the dielectric tensor $\overleftrightarrow {\mathbf{K}}$ in relativistic plasmas. The derivation of $\overleftrightarrow {\mathbf{K}}$ is explained by many plasma physics textbooks such as \citet{1992wapl.book.....S} and \citet{gurnett_bhattacharjee_2017}, and the dielectric tensor in relativistic plasmas is then
\begin{equation}\label{delectrictensor}
\begin{split}
    &\overleftrightarrow {\mathbf{K}}=\overleftrightarrow {\mathbf{1}}-\frac{\pi}{2}\sum_{j,n} \int\displaylimits_{0}^{\infty}\int\displaylimits_{-\infty}^{\infty}dp_{\parallel}dp_{\perp}\frac{   \omega_{pj}^2  p_{\perp}^2 \overleftrightarrow {\mathbf{S}}f_j }{\left[ p_\parallel-\frac{ m_j (\omega-n\Omega_{j})}{k_\parallel} \right]\omega^2 },
\end{split}
\end{equation}
where
\begin{equation}\label{Stensor}
\begin{split}
    \overleftrightarrow {\mathbf{S}}\equiv
\begin{bmatrix}
\epsilon_{xx}\hat G   & \epsilon_{xy}\hat G   & \epsilon_{xz}\hat H  \\
\epsilon_{yx}\hat G   & \epsilon_{yy}\hat G   & \epsilon_{yz}\hat H  \\
\epsilon_{zx}\hat G   & \epsilon_{zy}\hat G   & \epsilon_{zz}\hat H  \\
\end{bmatrix},
\end{split}
\end{equation}
\begin{equation}\label{Pixx}
\begin{split}
    &\epsilon_{xx}=\left(J_{n+1}e^{-i\phi_k}+J_{n-1}e^{i\phi_k}  \right)\! \left(J_{n+1}e^{i\phi_k}+J_{n-1}e^{-i\phi_k}  \right),
\end{split}
\end{equation}
\begin{equation}\label{Pixy}
\begin{split}
    &\epsilon_{xy}=-i\!\left(J_{n+1}e^{-i\phi_k}+J_{n-1}e^{i\phi_k}  \right)\!\left(J_{n+1}e^{i\phi_k}-J_{n-1}e^{-i\phi_k}  \right),
\end{split}
\end{equation}
\begin{equation}\label{Pixz}
\begin{split}
    &\epsilon_{xz}=2\frac{p_\parallel}{p_\perp}J_{n}\! \left(J_{n+1}e^{-i\phi_k}+J_{n-1}e^{i\phi_k}  \right),
\end{split}
\end{equation}
\begin{equation}\label{Piyx}
\begin{split}
    &\epsilon_{yx}=i\!\left(J_{n+1}e^{-i\phi_k}-J_{n-1}e^{i\phi_k}  \right)\!\left(J_{n+1}e^{i\phi_k}+J_{n-1}e^{-i\phi_k}  \right),
\end{split}
\end{equation}
\begin{equation}\label{Piyy}
\begin{split}
    &\epsilon_{yy}=\left(J_{n+1}e^{-i\phi_k}-J_{n-1}e^{i\phi_k}  \right)\!\left(J_{n+1}e^{i\phi_k}-J_{n-1}e^{-i\phi_k} \right),
\end{split}
\end{equation}
\begin{equation}\label{Piyz}
\begin{split}
    &\epsilon_{yz}=2i\frac{p_\parallel }{p_\perp}J_{n}\!\left(J_{n+1}e^{-i\phi_k}-J_{n-1}e^{i\phi_k}  \right),
\end{split}
\end{equation}
\begin{equation}\label{Pizx}
\begin{split}
    &\epsilon_{zx}=2\frac{p_\parallel }{p_\perp}J_{n}\!\left(J_{n+1}e^{i\phi_k}+J_{n-1}e^{-i\phi_k}  \right),
\end{split}
\end{equation}
\begin{equation}\label{Pizy}
\begin{split}
    &\epsilon_{zy}=-2i\frac{p_\parallel }{p_\perp}J_{n}\!\left(J_{n+1}e^{i\phi_k}-J_{n-1}e^{-i\phi_k}  \right),
\end{split}
\end{equation}
\begin{equation}\label{Pizz}
\begin{split}
    &\epsilon_{zz}=4\frac{p_\parallel^2 }{p_\perp^2}J_{n}^2,
\end{split}
\end{equation}
\begin{equation}
\begin{split}
    \hat G\equiv\left(\frac{m_j\omega}{k_\parallel}-p_\parallel\right)\frac{\partial }{\partial p_\perp} +p_\perp\frac{\partial }{\partial p_\parallel},
\end{split}
\end{equation}
\begin{equation}
\begin{split}
    \hat H\equiv\frac{m_jn\Omega_j}{k_\parallel}\frac{\partial }{\partial p_\perp} +\left(\frac{m_j\omega}{k_\parallel}-\frac{m_jn\Omega_j}{k_\parallel} \right)\frac{p_\perp}{p_\parallel}\frac{\partial }{\partial p_\parallel},
\end{split}
\end{equation}
and $\overleftrightarrow{\mathbf{1}}$ is the unit tensor. The subscript \textit{j} indicates the particle species. We denote the relativistic momentum coordinates perpendicular and parallel with respect to $\mathbf{B}_0$ as $p_\perp=m_j v_\perp$ and $p_\parallel=m_j v_\parallel$ where $m_j=\gamma m_{0j}$, $m_{0j}$ is the rest mass of particle  species \textit{j}, $v_\perp$ and $v_\parallel$ are the perpendicular and parallel velocity coordinates, and $\gamma=\sqrt{1+(p_\parallel^2+p_\perp^2)/m_{0j}^2c^2}$. The normalised and gyrotropic particle MDF of species \textit{j} is denoted as $f_{j}\equiv f_{j}(p_\perp,p_\parallel)$ so that $\int d^3\mathbf p f_j=1$. Note that, because $f_{j}$ is gyrotropic, we are free to select the value of $\phi_k$ in Eqs.~(\ref{Pixx}) to (\ref{Pizy}). 

The integer $n$ determines the order of the resonance, where $n=0$ corresponds to the Landau resonance and $n\neq 0$ corresponds to cyclotron resonances. The plasma frequency of species \textit{j} is defined as $\omega_{pj}^2=\omega_{0pj}^2/\gamma=4\pi n_{j}q_j^2/m_j$ where $q_j$ and $n_{j}$ are the charge and density of particle species \textit{j}. We denote the $n^{th}$-order Bessel function as $J_n=J_n(\rho_j)$ where $\rho_j \equiv k_\perp v_\perp/\Omega_{j}$. The cyclotron frequency of species \textit{j} is defined as $\Omega_j =\Omega_{0j}/\gamma= q_j B_0/m_jc$ and $c$ is the light speed. In this paper, we express that $\sum_{j}\sum_{n} \equiv \sum_{j,n}$.

The homogeneous wave equation is then expressed as 
\begin{equation}\label{asol}
\begin{split}
&\begin{bmatrix}
\!
K_{xx}\!-\!\frac{c^2\!(k_y^2+ k_{z}^2)}{\omega^2} \!\!\!&\!\!\! K_{xy}\!+\!\frac{c^2\!k_{x}k_{y}}{\omega^2} \!\!\!&\!\!\! K_{xz}\!+\!\frac{c^2\!k_{x}k_{z}}{\omega^2}\\[5pt]
K_{yx}\!+\!\frac{c^2\!k_{y}k_{x}}{\omega^2} \!\!\!&\!\!\! K_{yy}\!-\!\frac{c^2\!(k_x^2+k_z^2)}{\omega^2} \!\!\!&\!\!\! K_{yz}\!+\!\frac{c^2\!k_{y}k_{z}}{\omega^2}\\[5pt]
K_{zx}\!+\!\frac{c^2\!k_{z}k_{x}}{\omega^2} \!\!\!&\!\!\! K_{zy}\!+\!\frac{c^2\!k_{z}k_{y}}{\omega^2} \!\!\!&\!\!\! K_{zz}\!-\!\frac{c^2\!(k_x^2+k_y^2)}{\omega^2}
\!
\end{bmatrix}
\!\!
\begin{bmatrix}
E_k^x\\[5pt]
E_k^y\\[5pt]
E_k^z\\
\end{bmatrix}
\!=\!0.
\end{split}
\end{equation}
The condition that the non-trivial solution of Eq.~(\ref{asol}) exists is that the determinant of the $3\times3$ matrix in Eq.~(\ref{asol}) must be zero. This determinant corresponds to the plasma dispersion relation. Note that Eq.~(\ref{asol}) reduces to the homogeneous wave equation leading to the cold-plasma dispersion relation if we apply the cold-particle MDF, expressed in the plasma-rest frame as
\begin{equation}\label{coldvdf}
    f_j^{C}=\frac{1}{2\pi p_\perp}\delta(p_\perp)\delta(p_\parallel).
\end{equation}
We denote Dirac's $\delta$-function as $\delta$. The superscript $C$ indicates the cold particle population.

\subsection{Hermitian and Anti-Hermitian Part of the Dielectric Tensor}
To calculate the Hermitian part $\overleftrightarrow {\mathbf{K}}^h$, and anti-Hermitian part $\overleftrightarrow {\mathbf{K}}^a$, which appear in Eqs.~(\ref{waveenergy}) and (\ref{growthrate}), we find the real and imaginary part of the dielectric tensor Eq.~(\ref{delectrictensor}) based on the weak-growth rate approximation (i.e., $|\gamma_k|\ll |\omega_k|$). \footnote{\citet{rrefId0} discuss details about the real and imaginary parts of the dielectric tensor.} To do so, we use the Sokhotski-Plemelj relation for the $p_\parallel$-integral which faces the singularity problem in Eq.~(\ref{delectrictensor}). However, the $p_\parallel$-singularity in Eq.~(\ref{delectrictensor}), representing the resonant parallel momentum, is not yet in an explicit form because of the dependence of $\gamma$ on $p_\parallel$. Therefore, we rearrange the integrand of Eq.~(\ref{delectrictensor}) as
\begin{equation}\label{singularity}
\begin{split}
    &\frac{   \omega_{pj}^2  p_{\perp}^2 \overleftrightarrow {\mathbf{S}}f_j }{\left[ p_\parallel-\frac{ m_j (\omega-n\Omega_{j})}{k_\parallel} \right]\omega^2}\\
    &= \frac{   \omega_{pj}^2  p_{\perp}^2 \overleftrightarrow {\mathbf{S}}f_j\left [p_\parallel+\frac{ m_j (\omega+n\Omega_{j})}{k_\parallel}\right ]}{\left [ p_\parallel-\frac{ m_j (\omega-n\Omega_{j})}{k_\parallel}\right]\left [ p_\parallel+\frac{ m_j (\omega+n\Omega_{j})}{k_\parallel}\right]\omega^2}.
\end{split}
\end{equation}
We define the index of refraction in the direction parallel with respect to $\mathbf{B}_0$ as
\begin{equation}
    \eta_\parallel=\frac{c k_\parallel}{\omega},
\end{equation}
and, in the present paper, we only consider wave modes with 
\begin{equation}\label{indexcon}
    \eta_\parallel^2 > 1.
\end{equation}
Then, the $p_\parallel$-singularities of the right-hand side of Eq.~(\ref{singularity}) are
\begin{equation}\label{pres}
    p_{\parallel res}^{\pm}=-\frac{m_{0j}n\Omega_{0j}}{k_\parallel}+\gamma_{res}^{\pm}\frac{m_{0j}\omega}{k_\parallel},
\end{equation}
where %if $n \neq 0$ and $\eta_\parallel^2 \neq 1$,
\begin{equation}\label{cyclogammares}
\begin{split}
    \gamma_{res}^{\pm}=-\frac{n\Omega_{0j}}{\left(\eta_\parallel^2\!-\!1 \right)\omega}  \pm \eta_\parallel \! \sqrt{\frac{1}{\eta_\parallel^2\!-\!1} \!\left(1 \!+\!\frac{p_\perp^2}{c^2m_{0j}^2} \right)+\frac{n^2\Omega_{0j}^2}{\left(\eta_\parallel^2\!-\!1 \right)^2\!\omega^2}  }.
\end{split}
\end{equation}
%if $n \neq 0$ and $\eta_\parallel^2 = 1$,
%\begin{equation}\label{lightgammares}
%    \gamma_{res}^{\pm}=\frac{n\Omega_{0j}}{2\omega}+\frac{1}{2}\left(1+\frac{p_\perp^2}{c^2m_{0j}^2} \right)\frac{\omega}{n\Omega_{0j}},
%\end{equation}
%and if $n=0$ and $\eta_\parallel^2 > 1$,
%\begin{equation}\label{landaugammares}
%    \gamma_{res}^{\pm}=\pm |\eta_\parallel|  \sqrt{  \frac{1 +\frac{p_\perp^2}{c^2m_{0j}^2} }{\eta_\parallel^2-1}  }.
%\end{equation}
A careful analysis for Eq.~(\ref{pres}) with Eq.~(\ref{cyclogammares}) verifies that
\begin{equation}\label{a}
    \gamma (p_{\parallel res}^{\pm})=\pm\gamma_{res}^{\pm},
\end{equation}
and
\begin{equation}\label{a}
    p_{\parallel res}^{\pm} +\frac{m_{0j} n \Omega_{0j}}{k_\parallel} \mp \gamma(p_{\parallel res}^{\pm}) \frac{ m_{0j} \omega }{k_\parallel}=0.
\end{equation}

Based on the assumption that $|\gamma_k|\ll |\omega_k|$, the Sokhotski-Plemelj relation for the $p_\parallel$-integral, therefore, approximates Eq.~(\ref{delectrictensor}) as
\begin{equation}\label{Plemelj}
\begin{split}
    &\overleftrightarrow {\mathbf{K}}\approx\overleftrightarrow {\mathbf{1}}-\frac{\pi}{2}\sum_{j,n} \int\displaylimits_{0}^{\infty}\mint{-}\limits_{-\infty}^{\infty} dp_{\parallel}dp_{\perp}\frac{   \omega_{pj}^2  p_{\perp}^2 \overleftrightarrow {\mathbf{S}}f_j }{\left(p_\parallel-p_{\parallel res}^{+}\right)\omega_k^2\Lambda_{\alpha}}\\
    &-i \frac{k_\parallel}{|k_\parallel|}\frac{\pi^2}{2} \sum_{j,n}\int\displaylimits_{0}^{\infty}\int\displaylimits_{-\infty}^{\infty} dp_{\parallel}dp_{\perp}  \frac{   \omega_{pj}^2  p_{\perp}^2 \overleftrightarrow {\mathbf{S}}f_j}{ \omega_k^2\Lambda_{\beta}} \delta \! \left(p_\parallel-p_{\parallel res}^{+}\right),
\end{split}
\end{equation}
where
\begin{equation}
    \Lambda_{\alpha}=\left(1-\frac{1}{\eta_\parallel^2}\right)\frac{k_\parallel \left (p_\parallel-p_{\parallel res}^{-}\right)}{k_\parallel p_\parallel+m_j \left(\omega_k+n\Omega_{j}\right)},
\end{equation}
and
\begin{equation}
    \Lambda_{\beta}=1-\frac{p_\parallel}{\gamma m_{0j}c\eta_\parallel}.
\end{equation}
At the second term of the right-hand side of Eq.~(\ref{Plemelj}), the integral with a dash indicates the principal value integral, defined as
\begin{equation}\label{PV}
\begin{split}
    &\mint{-}\limits_{-\infty}^{\infty}dp_\parallel \equiv \lim_{\tau \rightarrow 0}\left(\int\displaylimits_{-\infty}^{p_{\parallel res}^{+}-\tau}dp_\parallel+\int\displaylimits_{p_{\parallel res}^{+}+\tau}^{\infty}dp_\parallel\right).
\end{split}
\end{equation}
Note that, after applying the Sokhotski-Plemelj relation, $\omega$ becomes $\omega_k$ in all following equations. By applying Eq.~(\ref{Plemelj}) to Eqs.~(\ref{hermitian}) and (\ref{antihermitian}), we then find the Hermitian part $\overleftrightarrow {\mathbf{K}}^h$ and anti-Hermitian part $\overleftrightarrow {\mathbf{K}}^a$, given as
\begin{equation}\label{finalhermitian}
\begin{split}
    \overleftrightarrow {\mathbf{K}}^h=\overleftrightarrow{\mathbf{1}}- \frac{\pi}{2}\sum_{j,n} \int\displaylimits_{0}^{\infty}\mint{-}\limits_{-\infty}^{\infty} dp_{\parallel}dp_{\perp}\frac{\omega_{pj}^2 p_{\perp}^2\overleftrightarrow {\mathbf{T}}f_j}{\left( p_\parallel-p_{\parallel res}^{+}\right)\omega_k^2 \Lambda_{\alpha}},
\end{split}
\end{equation}
and
\begin{equation}\label{finalantihermitian}
\begin{split}
    \overleftrightarrow {\mathbf{K}}^a=-i\frac{k_\parallel}{|k_\parallel|}\frac{\pi^2}{2}\sum_{j,n}\int\displaylimits_{0}^{\infty}&\int\displaylimits_{-\infty}^{\infty} dp_{\parallel}dp_{\perp} \\
    &\times\frac{ \omega_{pj}^2 p_{\perp}^2\overleftrightarrow {\mathbf{S}}f_j}{\omega_k^2 \Lambda_{\beta}} \delta\!\left(p_\parallel-p_{\parallel res}^{+}\right),
\end{split}
\end{equation}
where
\begin{equation}\label{Ttensor}
\begin{split}
    \overleftrightarrow {\mathbf{T}}\equiv
\begin{bmatrix}
\epsilon_{xx}\hat G   \!&\! \epsilon_{xy}\hat G   \!&\! \epsilon_{xz} \frac{\hat H+\hat G}{2} \\[5pt]
\epsilon_{yx}\hat G  \!&\! \epsilon_{yy} \hat G  \!&\! \epsilon_{yz}\frac{\hat H+\hat G}{2} \\[5pt]
\epsilon_{zx}\frac{\hat H+\hat G}{2}  \!&\! \epsilon_{zy}\frac{\hat H+\hat G}{2}  \!&\! \epsilon_{zz}\hat H \\
\end{bmatrix}.
\end{split}
\end{equation}
Note that $\hat G \equiv \hat H$ at $p_\parallel= p_{\parallel res}^{+}$. %We use Eqs.~(\ref{finalhermitian}) and (\ref{finalantihermitian}) to calculate Eqs.~(\ref{growthrate}) and (\ref{waveenergy}). 

\subsection{Energy Density and Growth Rate of Resonant Waves}

We calculate the energy density of the resonant wave Eq.~(\ref{waveenergy}) by using Eq.~(\ref{finalhermitian}). In calculating the second term of Eq.~(\ref{waveenergy}), we obtain
%\begin{equation}\label{pvrelation}
%\begin{split}
%    \frac{\partial}{\partial \omega_k}\Bigg[\int\displaylimits_{0}^{\infty}\! \mint{-}\limits_{-\infty}^{\infty}&dp_{\parallel} dp_{\perp} \frac{U_{\mu\nu}(p_\parallel)}{p_\parallel-p_{\parallel res}^{+}} \Bigg]\\
%    =\int\displaylimits_{0}^{\infty}\!\mint{-}\limits_{-\infty}^{\infty} &dp_{\parallel}dp_{\perp} \frac{1}{p_\parallel-p_{\parallel res}^{+}}\Bigg[ \frac{\partial U_{\mu\nu}(p_\parallel)}{\partial \omega_k}\\
%    &+ \frac{ U_{\mu\nu}(p_\parallel)-U_{\mu\nu}(p_{\parallel res}^{+}) }{p_\parallel-p_{\parallel res}^{+} }\frac{\partial p_{\parallel res}^{+}}{\partial \omega_k}\Bigg ],
%\end{split}    
%\end{equation}
\begin{equation}\label{pvrelation}
\begin{split}
    \frac{\partial}{\partial \omega_k}\left(\omega_k \overleftrightarrow {\mathbf{K}}^h\right)=\overleftrightarrow{\mathbf{1}}&- \frac{\pi}{2}\sum_{j,n} \int\displaylimits_{0}^{\infty}\mint{-}\limits_{-\infty}^{\infty} dp_{\parallel}dp_{\perp} \frac{\omega_{0pj}^2  p_{\perp}^2}{p_\parallel-p_{\parallel res}^{+}}\\
    &\times\left [ \left( \frac{ \overleftrightarrow {\mathbf{U}}-\overleftrightarrow {\mathbf{U}}_{res} }{p_\parallel-p_{\parallel res}^{+} }\right)\frac{\partial p_{\parallel res}^{+}}{\partial \omega_k}+\frac{\partial \overleftrightarrow {\mathbf{U}}}{\partial \omega_k}\right ],
\end{split}    
\end{equation}
where $\overleftrightarrow {\mathbf{U}}= \overleftrightarrow {\mathbf{T}}\!f_j/(\omega_k \gamma\Lambda_{\alpha})$ and $\overleftrightarrow {\mathbf{U}}_{res}\equiv \overleftrightarrow {\mathbf{U}} \big| _{p_\parallel= p_{ \parallel res}^{+}}$.
Eq.~(\ref{pvrelation}) is obtained through the definition of the principal value integral given as Eq.~(\ref{PV}), Leibniz integral rule, and the following relation
\begin{equation}
\begin{split}
    &\mint{-}\limits_{-\infty}^{\infty}dp_\parallel \frac{1}{\left(p_\parallel-p_{\parallel res}^{+}\right)^2}=\lim_{\tau \rightarrow 0}\frac{2}{\tau}.
\end{split}
\end{equation}
Then, by using Eq.~(\ref{pvrelation}), Eq.~(\ref{waveenergy}) is expressed as
\begin{equation}\label{finalenergy}
\begin{split}
    &W=W_{\text{Field}}+W_{\text{Particle}}.
\end{split}    
\end{equation}
The first term of Eq.~(\ref{finalenergy}) is the pure energy density of the wave fields, expressed as
\begin{equation}\label{fieldenergy}
    W_{\text{Field}}=\frac{1}{16\pi}\!\left(|\mathbf{{B}}_k|^2+|\mathbf{{E}}_k|^2 \right).
\end{equation}
The second term of Eq.~(\ref{finalenergy}) is the energy density of the wave fields associated with the kinetic energy of particles through Landau and cyclotron resonances, expressed as
\begin{equation}\label{finalenergyntotal}
    W_{\text{Particle}}=\sum_{j,n}W_{j}^n,
\end{equation}
where
\begin{equation}\label{finalenergyn}
\begin{split}
    &W_{j}^n=-\frac{1}{16} \int\displaylimits_{0}^{\infty}\mint{-}\limits_{-\infty}^{\infty} dp_{\parallel}dp_{\perp}\frac{\omega_{0pj}^2 p_{\perp}^2}{p_\parallel-p_{\parallel res}^{+}}\\
    &\times\left\{\frac{ \hat\Pi f_j-(\hat\Pi f_j)_{res} }{\left(p_\parallel-p_{ \parallel res}^{+}\right)\omega_k }\frac{\partial p_{\parallel res}^{+}}{\partial \omega_k}  +\frac{\partial }{\partial \omega_k}\left( \frac{\hat\Pi f_j}{\omega_k}\right)-\frac{\hat\Pi_\partial f_j}{\omega_k^2} \right\},
\end{split}    
\end{equation}
\begin{equation}
    \hat\Pi\equiv\frac{|\Psi|^2}{ \gamma\Lambda_{\alpha}}\hat G+\frac{2\textit{Re}\left[\Psi(\Psi^{z})^*\right]}{ \gamma\Lambda_{\alpha}}\left(\hat H-\hat G\right),
\end{equation}
\begin{equation}\label{hermitianpsi}
\begin{split}
\Psi 
\equiv  \mathit{{E}}_{k}^{R} e^{i\phi_k}  J_{n+1} +\mathit{{E}}_{k}^{L}& e^{-i\phi_k} J_{n-1}
+2\Psi^{z},
\end{split}
\end{equation}
\begin{equation}
    \Psi^{z}\equiv\frac{1}{\sqrt{2}}\frac{p_\parallel}{p_\perp} \mathit{{E}}_{k}^{z}J_{n},
\end{equation}
\begin{equation}
\begin{split}
    \hat{\Pi}_\partial\!\equiv\!\frac{2\textit{Re}\left[\Psi({\Psi}_\partial)^{*}\right]}{ \gamma\Lambda_{\alpha}}\hat G+\frac{2\textit{Re}\left[\Psi(\Psi_\partial^z)^{*}\!+\!\Psi_\partial (\Psi^z)^{*}\right]}{ \gamma\Lambda_{\alpha}}\!\left(\hat H-\hat G\right) ,
\end{split}
\end{equation}
\begin{equation}\label{hermitianprimepsi}
\begin{split}
\Psi_\partial
\equiv  \omega_k\frac{\partial \mathit{{E}}_{k}^{R}}{\partial \omega_k} e^{i\phi_k}  J_{n+1} +\omega_k\frac{\partial \mathit{{E}}_{k}^{L}}{\partial \omega_k}& e^{-i\phi_k} J_{n-1}+2\Psi_\partial^z,
\end{split}
\end{equation}
\begin{equation}
    {\Psi}_\partial^z\equiv\frac{\omega_k}{\sqrt{2}}\frac{p_\parallel}{p_\perp} \frac{\partial \mathit{{E}}_{k}^{z}}{\partial \omega_k}J_{n},
\end{equation}
and $(\hat\Pi f_j)_{res}\equiv \hat\Pi f_j \big| _{p_\parallel= p_{ \parallel res}^{+}}$. \textit{Re} indicates the real part of a complex function. We denote the right- and left-circularly polarised components of the electric field as $\mathit{{E}}_{k}^{R}\equiv(\mathit{{E}}_{k}^{x}-i\mathit{{E}}_{k}^{y})/\sqrt{2}$ and $\mathit{{E}}_{k}^{L}\equiv(\mathit{{E}}_{k}^{x}+i\mathit{{E}}_{k}^{y})/\sqrt{2}$. The longitudinal
component of the electric field is $\mathit{{E}}_{k}^{z}$. Note that, if the plasma is cold (i.e., if $f_j=\overline n_j^{C} f_j^C$ where $\overline n_j^{C}$ is the relative density of the cold particle population of species $j$ in total particles), Eq.~(\ref{finalenergyntotal}) reduces to
\begin{equation}\label{finalcoldenergy}
\begin{split}
    W_{\text{Particle}}^{C}=\frac{1}{16\pi}&\sum_{j} \overline n_j^{C}\omega_{0pj}^2\\
    &\times\left\{\frac{|\mathit{{E}}_{k}^{L}|^2}{\left(\omega_k-\Omega_{0j}\right)^2}+\frac{|\mathit{{E}}_{k}^{R}|^2}{\left(\omega_k+\Omega_{0j}\right)^2} +\frac{|\mathit{{E}}_{k}^{z}|^2}{\omega_k^2}\right\}.
\end{split}    
\end{equation}

Finally, by substituting Eq.~(\ref{finalantihermitian}) into Eq.~(\ref{growthrate}), the growth rate for a resonant wave in relativistic plasmas is expressed as
\begin{equation}\label{finalgrowthrate}
\begin{split}
    \gamma_k=\frac{k_\parallel}{|k_\parallel|}\sum_{j,n} \gamma_{k,j}^n,
\end{split}
\end{equation}
where
\begin{equation}\label{ngrowthrate}
\begin{split}
\gamma_{k,j}^n=\frac{\pi}{16W}\!\int\displaylimits_{0}^{\infty}\!\int\displaylimits_{-\infty}^{\infty}\! dp_{\parallel}dp_{\perp}\frac{ \omega_{pj}^2 p_{\perp}^2|\Psi|^2\hat G f_j}{\omega_k \Lambda_{\beta}}\delta\!\left(p_\parallel-p_{\parallel res}^{+}\right) .
\end{split}
\end{equation}
In the resonance with the wave mode of $\eta_\parallel^2 >1$, a non-relativistic limit is achievable by setting $c \rightarrow \infty $ in Eqs.~(\ref{finalenergyntotal}) and (\ref{finalgrowthrate}). By doing so, Eq.~(\ref{finalgrowthrate}) reduces to the growth rate given by \citet{kennel_wong_1967}. Note that the only approximation, which the derivation of Eqs.~(\ref{finalenergyntotal}) and (\ref{finalgrowthrate}) is based on, is that $|\gamma_k|\ll |\omega_k|$.

\section{Correlation between the Wave Energy Density and Growth Rate} \label{3}

We demonstrate that Eq.~(\ref{finalenergyntotal}) for the wave energy density associated with the particle's kinetic energy through resonances and Eq.~(\ref{finalgrowthrate}) for the wave growth rate are correlated across wavenumber. Under the action of an instability, the unstable wave scatters particles, and absorbs their kinetic energy. This means that, within a specific wavenumber range where Eq.~(\ref{finalgrowthrate}) becomes positive through a specific resonance, Eq.~(\ref{finalenergyntotal}) must show an explicit energy density increase through the same resonance. To demonstrate this, we numerically analyse an instability that operates in the near-Earth high-energy environment of the radiation belts. We choose here the loss-cone driven instability since a lot of \textit{in-situ} data is provided, and this instability occurs through waves both parallel and obliquely propagating with respect to $\mathbf{B}_0$ in relativistic plasmas \citep{https://doi.org/10.1029/JZ071i001p00001}.

%indicate that, during times of high values of electron flux in Earth's radiation belts, a flux-limiting process acts \citep{https://doi.org/10.1029/2021GL095779}, possibly mediated by the loss-cone driven instability as suggested by \citet{https://doi.org/10.1029/JZ071i001p00001}.

In the present paper, we define the $n$ resonance as the
contribution to the summation in Eqs.~(\ref{finalenergyntotal}) and (\ref{finalgrowthrate}) with only integer $n$. We only consider the $n=+1$,
$-1$, and $0$ resonances, ignoring higher-$n$ resonances due to
their negligible contributions. A more complete study, including the time evolution of relativistic electrons in the momentum space through the loss-cone driven instability, requires an additional analysis with the quasi-linear theory in relativistic plasmas \citep{Jeong_2020}. However, this is beyond the scope of the present paper. This paper only focuses on the demonstration for the correlation between Eqs.~(\ref{finalenergyntotal}) and (\ref{finalgrowthrate}), which are in the same structure.

%numerically investigate the loss-cone driven instability, as a practical example, in which the whistler-mode waves scatter the relativistic electrons into the loss cone in the radiation belt \citep{https://doi.org/10.1029/JZ071i001p00001,lyons_thorne_kennel_1971}.

\subsection{Background Plasma for the Loss-Cone Driven Instability}

Earth’s inner magnetosphere includes a relatively cold and dense ($\sim 1$~eV) electron population originating from the Earth’s ionosphere, alongside more sparse, but equally important, energetic electron populations ($>1$~keV) which are sourced from the magnetotail plasma sheet and further processed by wave-particle interactions in the outer radiation belt \citep{Borovsky2018TheEM}. These energetic electron populations have been studied statistically from geosynchronous orbit \citep{https://doi.org/10.1029/2009JA014183}, demonstrating that there are typically two energetic electron populations: a hotter population with a number density of $\sim 5 \times 10^{-4}$~cm$^{-3}$ and a temperature of $\sim150$~keV, and a less-hot population with a number density of $\sim 10^{-2}$~cm$^{-3}$ and a temperature of $\sim 30$~keV. If these energetic electron populations exhibit the loss-cone MDF, they are unstable to whistler-mode wave in the inner magnetosphere \citep{https://doi.org/10.1002/2017JA024399,articleLi}. 

\subsection{Overall Scheme for Numerical Calculations} \label{3.1}

In our numerical calculations for Eqs.~(\ref{finalenergyntotal}) and (\ref{finalgrowthrate}), we first assume that dominant cold electron and proton populations, and a minor relativistic electron population exist in the radiation belt. In the plasma-rest frame, for the cold populations, we use Eq.~(\ref{coldvdf}), while, for the relativistic electron population, we use the loss-cone MDF typically expressed in the plasma-rest frame as \citep{doi:10.1063/1.872932}
\begin{equation}\label{lossconeMDF}
    f_e^{R}=\frac{2\Gamma\!\left(\frac{q+3}{2} \right)\Gamma\!\left(s+1 \right)}{\pi^2 p_0^3\Gamma\!\left(\frac{q+2}{2} \right)\Gamma\!\left(s-\frac{1}{2} \right)}\left[\frac{p_\perp^q}{\left(1+\frac{p_\perp^2+p_\parallel^2}{p_0^2} \right)^{s+1}\!\!\sqrt{p_\perp^2+p_\parallel^2}^q}\right],
\end{equation}
where
\begin{equation}
    p_0=\left \{\frac{\left(q+3 \right)\left(2s-3 \right)}{3} \right \}^{\frac{1}{2}}m_{0e}v_{th,e},
\end{equation}
$v_{th,e}\equiv \sqrt{2k_BT_e^R/m_{0e}}$ and $\Gamma(x)$ is the $\Gamma$-function. The superscript $R$ indicates the relativistic particle population. The free parameter $q$ and $s$ determine the anisotropy and steepness of the loss-cone MDF, $T_{e}^R$ is the temperature of the relativistic electron population, and $k_B$ is the Boltzmann constant. For our calculation, we then express the proton and electron MDFs of a quasi-neutral plasma as
\begin{equation}\label{totalp}
    f_{p}=f_p^{C},
\end{equation}
and
\begin{equation}\label{totale}
    f_{e}=\overline n_e^{C}f_e^{C}+\overline n_e^{R}f_e^{R},
\end{equation}
where $\overline n_e^{C}$ and $\overline n_e^{R}$ indicate the relative density of the cold and relativistic electron population in total electrons, and $\overline n_e^{C}+\overline n_e^{R}=1$. The subscript $p$ and $e$ indicate the proton and electron. 

\begin{figure}
  \centering
  \includegraphics[width=\columnwidth]{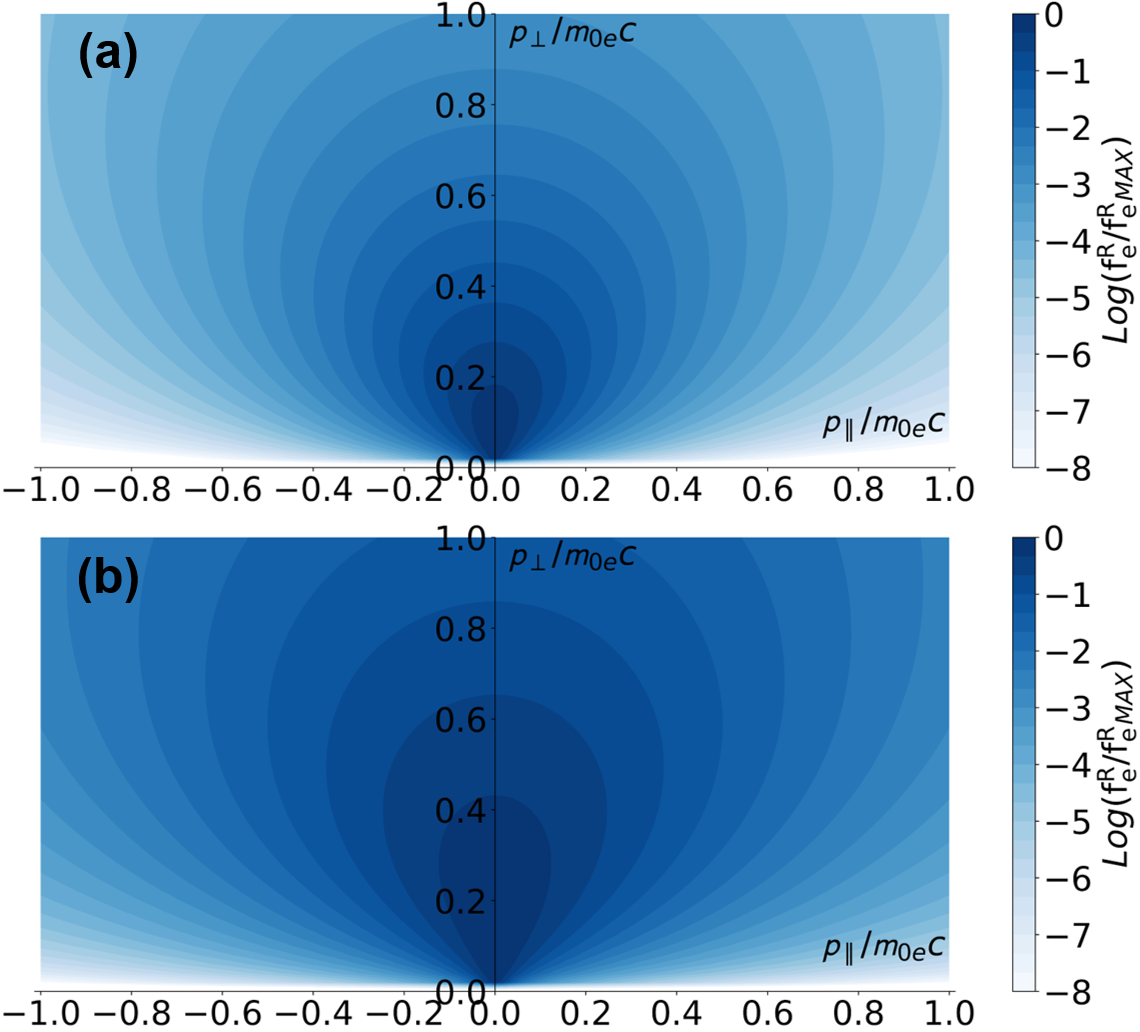}
  \caption{Unstable loss-cone MDFs for the relativistic electron populations, plotted through Eq.~(\ref{lossconeMDF}) with our plasma parameters of (a) soft and (b) hard populations. The values of the MDFs are normalised to their maximum values, respectively. We apply these MDFs to calculate Eqs.~(\ref{finalenergyntotal}) and (\ref{finalgrowthrate}).}
  \label{initiallossconeMDF}
\end{figure}

In our calculation, we set the plasma parameters, which
are representative in the radiation belt of the Earth, notwithstanding the wide range of natural variation \citep{https://doi.org/10.1029/2010JA016151,https://doi.org/10.1029/2018JA026401}. We calculate two different cases for the relativistic electron population, set based on the \textit{in-situ} data \citep{https://doi.org/10.1029/2009JA014183,https://doi.org/10.1002/2014GL060707} and defined as 
\begin{enumdescript}
   \item Soft population: $T_{e}^R=25$~keV and $\overline n_e^{R}=0.05$, 
   \item Hard population: $T_{e}^R=140$~keV and $\overline n_e^{R}=0.0025$.
\end{enumdescript}
Fig.~\ref{initiallossconeMDF} illustrates the normalised $f_e^R$, which are the initially unstable loss-cone MDFs for the relativistic electron populations and plotted through Eq.~(\ref{lossconeMDF}) with our plasma parameters of the (a) soft and (b) hard populations. In Eq.~(\ref{lossconeMDF}), we arbitrarily set to $q=4$ and $s=2$ for our calculation due to the lack of \textit{in-situ} data, which need to be determined by fitting with \textit{in-situ} data at future studies. We set $\omega_{0pe}=3|\Omega_{0e}|$ \citep{https://doi.org/10.1002/jgra.50594}.

To calculate Eqs.~(\ref{finalenergyntotal}) and (\ref{finalgrowthrate}), we preferentially find the frequency and polarisation properties of the whistler-mode wave as a function of $k_\parallel$ by numerically solving a dispersion relation from Eq.~(\ref{asol}). We confirm that the effect of our relativistic electron populations on the solution of a full relativistic-plasma dispersion relation is very subtle due to its small density (i.e., $5\%$ of soft population or $0.25\%$ of hard population in total electrons) compared to the cold plasma. Thus, to reduce the computational amount and time in finding its solution, we only apply $\overline n_e^{C}f_e^{C}$ of Eq.~(\ref{totale}) to Eq.~(\ref{asol}), which means that we numerically solve the cold-plasma dispersion relation in our calculation. \footnote{In case that the effect of the relativistic electron population is not negligible, a full relativistic-plasma dispersion relation must be solved. To do so, the ALPS (Arbitrary Linear Plasma Solver) is an appropriate tool, which is an advanced and parallelised code numerically solving a full dispersion relation in relativistic regime \citep{verscharen_klein_chandran_stevens_salem_bale_2018}.} %Because $|\gamma_k|\ll |\omega_k|$, we set $\gamma_k=0$ in Eq.~(\ref{asol}) for the simplicity of our numerical calculations when finding $\omega_k\equiv \omega_k(k_\parallel)$. 

Without loss of generality, we set $\omega_k>0$ in this calculation. We set $\phi_k=0^\circ$ because Eqs.~(\ref{totalp}) and (\ref{totale}) are gyrotropic, and denote the angle of the wave propagation with respect to $\mathbf{B}_0$ as $\theta$ (i.e., $k_x=k_\perp=k_\parallel \tan\theta$ and $k_y=0$). We choose our coordinate system so that the whistler-mode waves with an arbitrary angle $\theta$ have $k_\parallel>0$. We then find $\omega_k\equiv \omega_k(k_\parallel)$, which makes the determinant of the $3\times3$ matrix in Eq.~(\ref{asol}) zero, in cold-plasma regime and with a given $\theta$. In doing so, we use the \textit{fsolve} function in our Python code, which finds the root of the non-linear equation. Once we find $\omega_k\equiv\omega_k(k_\parallel)$, we use it to find the polarisation properties of the whistler-mode wave from Eq.~(\ref{asol}) as the ratios $|\mathit{{E}}_{k}^{y}|/|\mathit{{E}}_{k}^{x}|$ and $|\mathit{{E}}_{k}^{z}|/|\mathit{{E}}_{k}^{x}|$, and as a function of $k_\parallel$. In this way, because we are at liberty to set $E_k^x$, we set $E_k^x$ as a constant.

By using the profiles of $\omega_k$, $|\mathit{{E}}_{k}^{y}|/|\mathit{{E}}_{k}^{x}|$ and $|\mathit{{E}}_{k}^{z}|/|\mathit{{E}}_{k}^{x}|$, we calculate Eqs.~(\ref{finalenergyntotal}) and (\ref{finalgrowthrate}) as a function of $k_\parallel$ based on plasma parameters of soft and hard populations, respectively. We numerically evaluate the principal value integral in Eq.~(\ref{finalenergyntotal}), defined as Eq.~(\ref{PV}), through the following relation,
\begin{equation}\label{amari}
\begin{split}
    \mint{-}\limits_{a}^{b} dX  \frac{h(X)}{X-X_s} = \int\displaylimits_{a}^{b} dX \frac{h(X)-h(X_s)}{X-X_s}+h(X_s)\ln \frac{b-X_s}{X_s-a},
\end{split}
\end{equation}
where
$h(X)$ is an arbitrary function of the variable $X$, and $X_s$ indicates a singularity. We assume that the integration of the first term on the right-hand side of Eq.~(\ref{amari}) converges. We normalise $p_\perp$ and $p_\parallel$ by $m_{0e}c$, and frequency by $|\Omega_{0e}|$. For numerical reasons in our calculations, we set the finite ranges of $p_\perp/m_{0e}c$ and $p_\parallel/m_{0e}c$ as $0.01<p_\perp/m_{0e}c\leq3.8$ and $-3.8\leq p_\parallel/m_{0e}c\leq3.8$ for integrals in Eqs.~(\ref{finalenergyntotal}) and (\ref{finalgrowthrate}), confirming that these ranges are wide enough to make the integration converge. We numerically calculate the $p_\parallel$- and $p_\perp$-integration based on the Simpson's $3/8$ rule. We set the range of $ck_\parallel/|\Omega_{0e}|$ as $0.3<ck_\parallel/|\Omega_{0e}|\leq10$, and the step size of $ck_\parallel/|\Omega_{0e}|$ is $0.065$ in our calculation.

\begin{figure}
  \centering
  \includegraphics[width=\columnwidth]{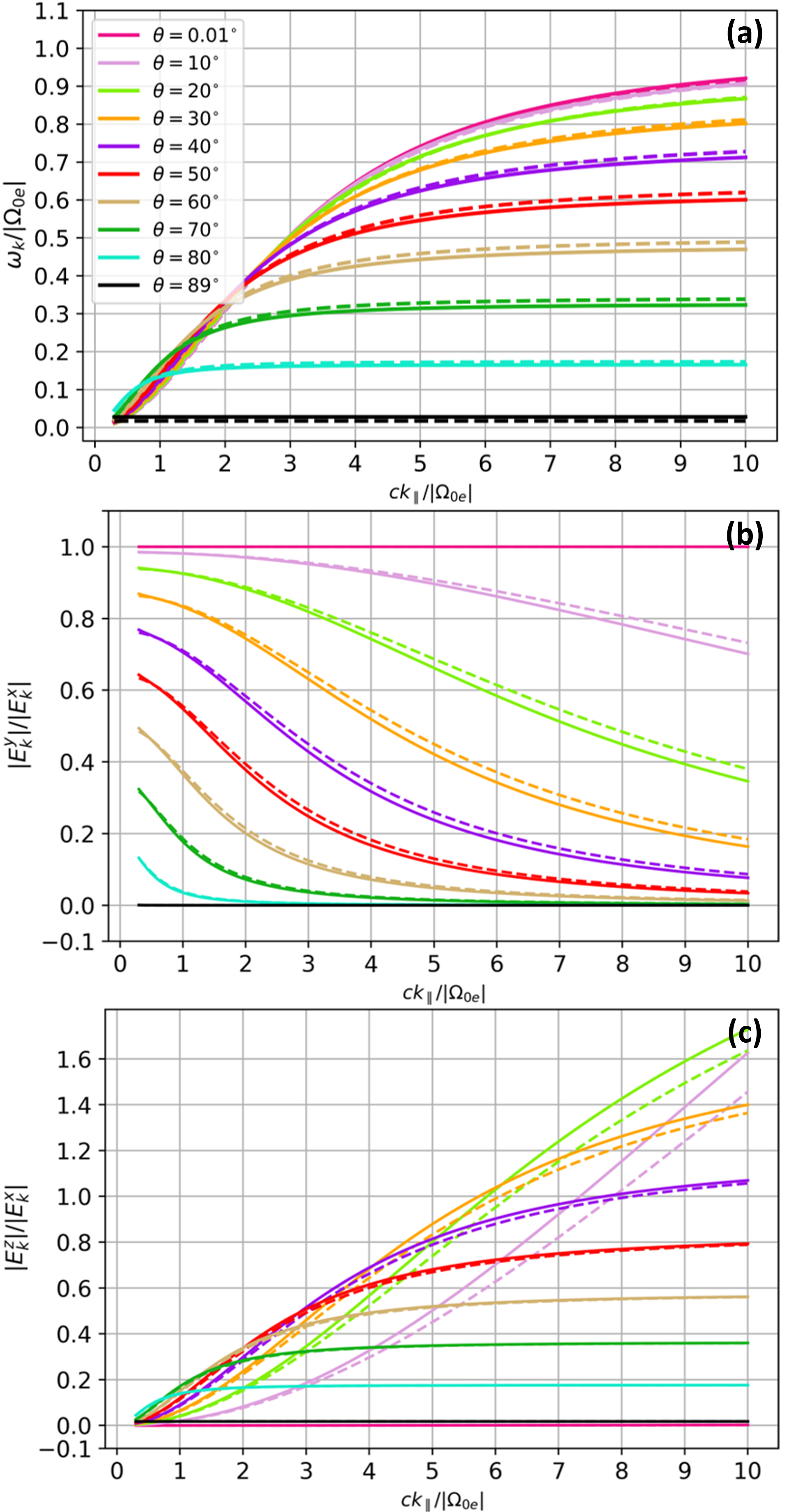}
  \caption{Profiles of (a) $\omega_k$, (b) $|\mathit{{E}}_{k}^{y}|/|\mathit{{E}}_{k}^{x}|$, and (c) $|\mathit{{E}}_{k}^{z}|/|\mathit{{E}}_{k}^{x}|$ as a function of $k_\parallel$ with different propagation angles $\theta$, ranging from $\theta=0.01^\circ$ to $\theta=89^\circ$. Solid curves are the numerical results from the cold-plasma dispersion relation and dashed curves with the same color scheme for $\theta$ are the analytical results from Eqs.~(\ref{analyticfre}) to (\ref{analyticEz}). These two results are consistent with each other in each figure. We use the numerical result to calculate Eqs.~(\ref{finalenergyntotal}) and (\ref{finalgrowthrate}).}
  \label{results}
\end{figure}

\subsection{Numerical Demonstration for the Correlation} \label{3.2}

Fig.~\ref{results} shows the profiles of $\omega_k$, $|\mathit{{E}}_{k}^{y}|/|\mathit{{E}}_{k}^{x}|$, and $|\mathit{{E}}_{k}^{z}|/|\mathit{{E}}_{k}^{x}|$ as a function of $k_\parallel$ with different propagation angles, ranging from $\theta=0.01^\circ$ to $\theta=89^\circ$. In each figure, the solid curves correspond to the numerical results from the cold-plasma dispersion relation. With the assumptions that the proton contribution is neglected, $\omega_k^2 \ll \omega_{0pe}^2$ and $\Omega_{0e}^2 \ll \omega_{0pe}^2$, the cold-plasma dispersion relation also provides the analytic expressions for $\omega_k$, $|\mathit{{E}}_{k}^{y}|/|\mathit{{E}}_{k}^{x}|$, and $|\mathit{{E}}_{k}^{z}|/|\mathit{{E}}_{k}^{x}|$ of the whistler-mode wave as \citep{1992wapl.book.....S}
\begin{equation}\label{analyticfre}
    \omega_k \approx \frac{c^2k_\parallel^2|\Omega_{0e}| \cos\theta}{c^2k_\parallel^2+\omega_{0pe}^2\cos^2\theta},
\end{equation}
\begin{equation}\label{analyticEy}
    \frac{|\mathit{{E}}_{k}^{y}|}{|\mathit{{E}}_{k}^{x}|} \approx \frac{\omega_{0pe}^2 |\Omega_{0e}|  \cos^2\theta \omega_k}{c^2k_\parallel^2 |\Omega_{0e}|^2-\left(\omega_{0pe}^2\cos^2\theta + c^2k_\parallel^2 \right)\omega_k^2},
\end{equation}
and
\begin{equation}\label{analyticEz}
    \frac{|\mathit{{E}}_{k}^{z}|}{|\mathit{{E}}_{k}^{x}|} \approx \frac{c^2k_\parallel^2 \tan\theta}{\omega_{0pe}^2+c^2k_\parallel^2\tan^2\theta}.
\end{equation}
The dashed curves in Fig.~\ref{results} with the $\theta$-colour scheme same as the $\theta$-colour scheme of solid curves correspond to the analytical results from Eqs.~(\ref{analyticfre}) to (\ref{analyticEz}). The numerical results for $\omega_k$, $|\mathit{{E}}_{k}^{y}|/|\mathit{{E}}_{k}^{x}|$ and $|\mathit{{E}}_{k}^{z}|/|\mathit{{E}}_{k}^{x}|$ closely follow the analytical results. This verifies that our numerical solutions of the cold-plasma dispersion relation are correct. 

\begin{figure}
  \centering
  \includegraphics[width=\columnwidth]{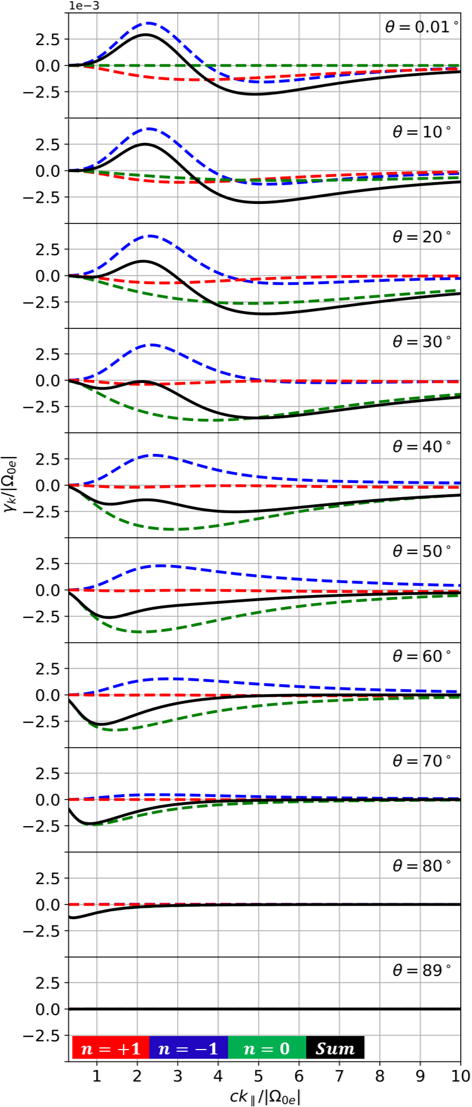}
  \caption{Profiles of $\gamma_k$ as a function of $k_\parallel$ with different propagation angles based on the soft population of relativistic electrons, ranging from $\theta=0.01^\circ$ to $\theta=89^\circ$. The solid black, and dashed red, blue, and green curves correspond to $\gamma_k$, $\gamma_{k,e}^{n=+1}$, $\gamma_{k,e}^{n=-1}$ and $\gamma_{k,e}^{n=0}$. The profiles show that the $n=-1$ resonance drives the instability, overcoming the $n=+1$ and $n=0$ resonances within a specific $k_\parallel$-range in the $\theta$-range from $\theta=0.01^\circ$ to $\theta=30^\circ$.}
  \label{growthrateresult}
\end{figure}

By using our numerical results for $\omega_k$, $|\mathit{{E}}_{k}^{y}|/|\mathit{{E}}_{k}^{x}|$, and $|\mathit{{E}}_{k}^{z}|/|\mathit{{E}}_{k}^{x}|$, we calculate Eq.~(\ref{finalgrowthrate}) as a function of $k_\parallel$ with different $\theta$ based on Eq.~(\ref{totale}). Note that the cold-particle MDF provides no contribution to Eq.~(\ref{finalgrowthrate}), which means that Eq.~(\ref{finalgrowthrate}) is only estimated at $f_e=\overline n_e^{R}f_e^{R}$, illustrated in Fig.~\ref{initiallossconeMDF}. Fig.~\ref{growthrateresult} shows the profiles of $\gamma_k$, $\gamma_{k,e}^{n=+1}$, $\gamma_{k,e}^{n=-1}$ and $\gamma_{k,e}^{n=0}$ estimated at $f_e=\overline n_e^{R}f_e^{R}$ for the soft population of relativistic electrons, ranging from $\theta=0.01^\circ$ to $\theta=89^\circ$. In each plot, the solid black curve, and dashed red, blue, and green curves correspond to $\gamma_k$, $\gamma_{k,e}^{n=+1}$, $\gamma_{k,e}^{n=-1}$ and $\gamma_{k,e}^{n=0}$, respectively. %The solid curves correspond to the relativistic situation while the dashed curves with the same color scheme correspond to the non-relativistic situation.
In Fig.~\ref{growthrateresult}, $\gamma_{k,e}^{n=-1}$ only contributes to the wave instability (i.e., the positive contribution to $\gamma_k$). On the other hand, $\gamma_{k,e}^{n=+1}$ and $\gamma_{k,e}^{n=0}$ contribute to the wave damping (i.e., the negative contribution to $\gamma_k$), and counteract the driving of the wave instability. Therefore, the $n=-1$ resonance must dominate over the $n=+1$ and $n=0$ resonances, so that the loss-cone driven instability occurs.
When $\theta=0^\circ$, the loss-cone driven instability is the most efficient. But, when increasing $\theta$, it becomes inefficient due to an increase of the wave damping through the $n=0$ resonance, and, from around $\theta=30^\circ$, the loss-cone driven instability disappears.

\begin{figure}
  \centering
  \includegraphics[width=\columnwidth]{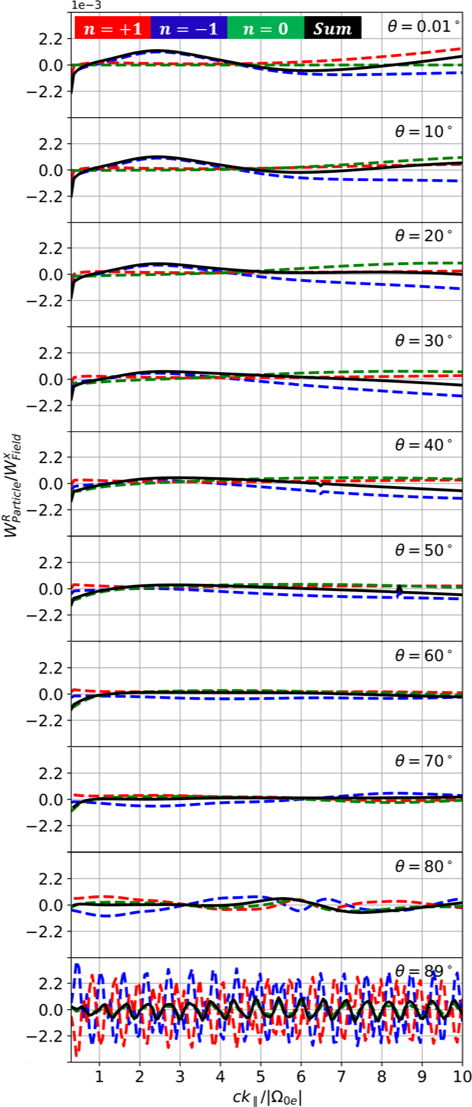}
  \caption{Profiles of $W_{\text{Particle}}^{R}$ as a function of $k_\parallel$ with different propagation angles based on the soft population  of relativistic electrons, ranging from $\theta=0.01^\circ$ to $\theta=89^\circ$, and we define that $W_{\text{Field}}^x=|E_k^x|^2/16\pi$. The solid black, and dashed red, blue and green curves correspond to $W_{\text{Particle}}^{R}$, $W_{e}^{R,n=+1}$, $W_{e}^{R,n=-1}$ and $W_{e}^{R,n=0}$. The profiles show that $W_{\text{Particle}}^{R}$ presents an explicit increase through the $n=-1$ resonance from $\theta=0.01^\circ$ to $\theta=30^\circ$, in the $k_\parallel$-range where $\gamma_k$ in Fig.~\ref{growthrateresult} is positive and the instability is driven.}
  \label{energyresult}
\end{figure}

We also calculate the energy density of the resonant whistler-mode wave Eq.~(\ref{finalenergy}) as a function of $k_\parallel$ with different $\theta$ when calculating $\gamma_k$, given as
\begin{equation}\label{rwenergy}
\begin{split}
    &W=W_{\text{Field}}+W_{\text{Particle}}^{C}+W_{\text{Particle}}^{R},
\end{split}    
\end{equation}
where 
\begin{equation}\label{rwenergyhot}
    W_{\text{Particle}}^{R}=\sum_{n}W_{e}^{R,n}=\sum_{n}W_{e}^{n}\Big |_{f_e=\overline n_e^{R}f_e^{R}}.
\end{equation}
Eq.~(\ref{rwenergyhot}) describes the energy density of whistler-mode waves associated with the kinetic energy of relativistic electrons through Landau and cyclotron resonances. Fig.~\ref{energyresult} shows the profiles of $W_{\text{Particle}}^{R}$, $W_{e}^{R,n=+1}$, $W_{e}^{R,n=-1}$ and $W_{e}^{R,n=0}$ estimated at $f_e=\overline n_e^{R}f_e^{R}$ for the soft population of relativistic electrons, ranging from $\theta=0.01^\circ$ to $\theta=89^\circ$. We define that $W_{\text{Field}}^x=|E_k^x|^2/16\pi$. In each plot, the solid black curve, and dashed red, blue, and green curves correspond to $W_{\text{Particle}}^{R}$, $W_{e}^{R,n=+1}$, $W_{e}^{R,n=-1}$ and $W_{e}^{R,n=0}$, respectively. From $\theta=0.01^\circ$ to $\theta=30^\circ$, $W_{\text{Particle}}^{R}$ has an explicit increase of the wave energy density due to the $n=-1$ resonance, roughly in the unstable $k_\parallel$-range where $\gamma_k$ in Fig.~\ref{growthrateresult} is positive. This explicit increase of the wave energy density becomes weaker when $\theta$ increases, and nearly disappears at $\theta=30^\circ$ where the loss-cone driven instability disappear. 

These behaviours of $W_{\text{Particle}}^{R}$ shown in Fig.~\ref{energyresult} are consistent with behaviours of $\gamma_k$ shown in Fig.~\ref{growthrateresult} in the perspective of the wave growth. From $\theta=30^\circ$, $W_{\text{Particle}}^{R}$ closely stays at zero on average because $W_{\text{Particle}}^{R,n=-1}$ counteracts to $W_{\text{Particle}}^{R,n=+1}$ and $W_{\text{Particle}}^{R,n=0}$, and cancel with each other.

\begin{figure}
  \centering
  \includegraphics[width=\columnwidth]{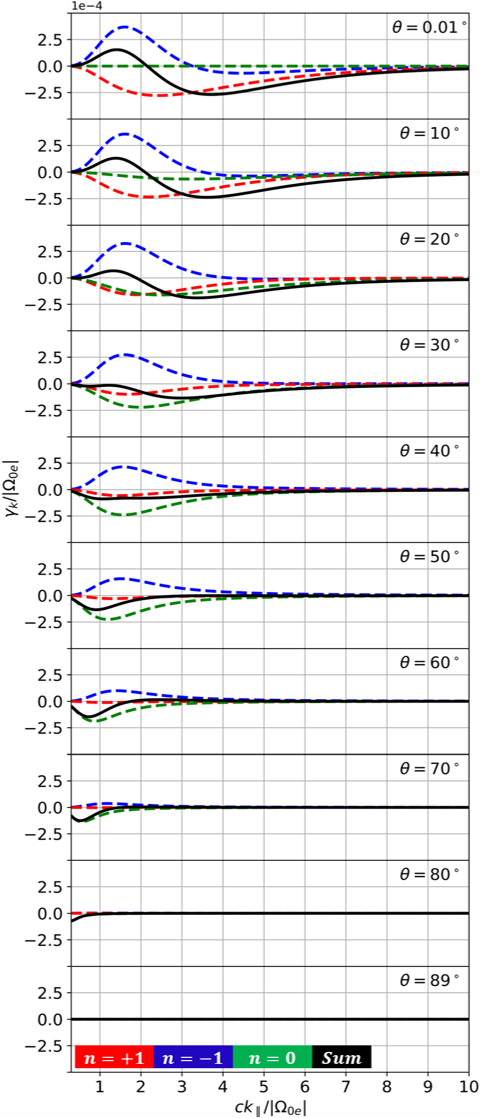}
  \caption{Profiles of $\gamma_k$ as a function of $k_\parallel$ with different propagation angles based on the hard population of relativistic electrons, ranging from $\theta=0.01^\circ$ to $\theta=89^\circ$. The solid black, and dashed red, blue, and green curves correspond to $\gamma_k$, $\gamma_{k,e}^{n=+1}$, $\gamma_{k,e}^{n=-1}$ and $\gamma_{k,e}^{n=0}$. The profiles show that the $n=-1$ resonance drives the instability, overcoming the $n=+1$ and $n=0$ resonances within a specific $k_\parallel$-range in the $\theta$-range from $\theta=0.01^\circ$ to $\theta=30^\circ$.}
  \label{growthrateresulthigh}
\end{figure}

\begin{figure}
  \centering
  \includegraphics[width=\columnwidth]{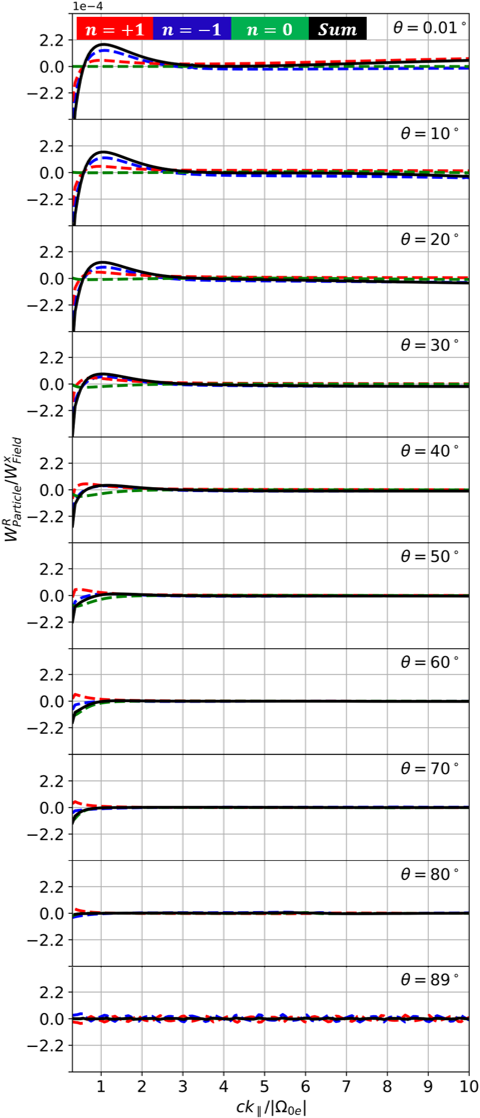}
  \caption{Profiles of $W_{\text{Particle}}^{R}$ as a function of $k_\parallel$ with different propagation angles based on the hard population of relativistic electrons, ranging from $\theta=0.01^\circ$ to $\theta=89^\circ$. The solid black, and dashed red, blue and green curves correspond to $W_{\text{Particle}}^{R}$, $W_{e}^{R,n=+1}$, $W_{e}^{R,n=-1}$ and $W_{e}^{R,n=0}$. The profiles show that $W_{\text{Particle}}^{R}$ presents an explicit increase through the $n=-1$ resonance from $\theta=0.01^\circ$ to $\theta=30^\circ$, in the $k_\parallel$-range where $\gamma_k$ in Fig.~\ref{growthrateresulthigh} is positive and the instability is driven.}
  \label{energyresulthigh}
\end{figure}

Fig.~\ref{growthrateresulthigh} and Fig.~\ref{energyresulthigh} show the profiles of $\gamma_k$, $\gamma_{k,e}^{n=+1}$, $\gamma_{k,e}^{n=-1}$ and $\gamma_{k,e}^{n=0}$, and the profiles of $W_{\text{Particle}}^{R}$, $W_{e}^{R,n=+1}$, $W_{e}^{R,n=-1}$ and $W_{e}^{R,n=0}$ estimated at $f_e=\overline n_e^{R}f_e^{R}$ for the hard population of relativistic electrons, ranging from $\theta=0.01^\circ$ to $\theta=89^\circ$. Even though the loss-cone driven instability occurs at a different $k_\parallel$-range from the case of the soft population, the general patterns from the hard populations are same as the patterns from the soft population. In Fig.~\ref{growthrateresulthigh}, $\gamma_{k,e}^{n=-1}$ only contributes to the wave instability, and dominates over $\gamma_{k,e}^{n=+1}$ and $\gamma_{k,e}^{n=0}$ within a specific $k_\parallel$-range from $\theta=0.01^\circ$ to $\theta=30^\circ$. In this $\theta$-range, $W_{\text{Particle}}^{R}$ in Fig.~\ref{energyresulthigh} has an explicit increase of the wave energy density due to the $n=-1$ resonance, roughly in the unstable $k_\parallel$-range where $\gamma_k$ in Fig.~\ref{growthrateresulthigh} is positive. Therefore, in the perspective of the wave growth, the behaviours of $W_{\text{Particle}}^{R}$ shown in Fig.~\ref{energyresulthigh} are also consistent with behaviours of $\gamma_k$ shown in Fig.~\ref{growthrateresulthigh}.

%We show that the relativistic effect makes the maximum growth rate from the $n=-1$ resonance weaker. In addition, the relativistic effect makes the damping rate from the $n=+1$ resonance weaker, while the relativistic effect makes the damping rate from the $n=0$ resonance stronger in the range of $k_\parallel$ where the instability occurs.

%The relativistic effect makes the maximum growth rate from the $n=-1$ resonance weaker. In the range of $k_\parallel$ where the instability occurs, the relativistic effect makes the damping rate from the $n=+1$ resonance weaker, but make the damping rate from the $n=0$ resonance stronger.


\section{DISCUSSION AND CONCLUSIONS} \label{4}

The kinetic linear theory has proven to explain many wave instabilities in space and astrophysical plasmas \citep{weirefId0,Verscharen_2013,2017RvMPP...1....4Y,Vasko_2019,Verscharen_2019,2021ApJ...916L...4S,Jiang_2022}. However, although the linear theory has been practically used for various instabilities, the basis for linear theory itself must be further developed to more deeply understand wave instabilities.

In this paper, we derive the analytic expressions for the energy density, Eq.~(\ref{finalenergy}), and growth rate, Eq.~(\ref{finalgrowthrate}), of an arbitrary resonant wave with an arbitrary propagation angle with respect to $\mathbf{B}_0$ in relativistic plasmas. We calculate the Hermitian and anti-Hermitian parts of the relativistic-plasma dielectric tensor, and complete our analytic expressions. Therefore, we explicitly formulate Eq.~(\ref{waveenergy}), which had previously been left unclear, and analytically express the energy density of the wave fields associated with the kinetic energy of particles through Landau and cyclotron resonances, given by Eq.~(\ref{finalenergyntotal}). In addition, we generalise the analytic expression for the growth rate given by \citet{kennel_wong_1967}. 

Under the action of an instability, the unstable wave absorbs the kinetic energy of resonant particles. Thus, if Eq.~(\ref{finalgrowthrate}) becomes positive through a specific resonance within a specific wavenumber range where a wave instability is driven, Eq.~(\ref{finalenergyntotal}) must present an explicit increase of the wave energy density through the same resonance within the same wavenumber range. The present paper numerically demonstrates this correlation between Eqs.~(\ref{finalenergyntotal}) and (\ref{finalgrowthrate}) by analysing the loss-cone driven instability in the background plasma of Earth's radiation belt, as a specific example. We analyse two different cases for the relativistic-electron population in the loss-cone MDF, which causes the unstable whistler-mode waves.

Fig.~\ref{growthrateresult} and Fig.~\ref{growthrateresulthigh}, corresponding to results from soft and hard populations of relativistic electrons respectively, illustrate Eq.~(\ref{finalgrowthrate}) (i.e., illustrate the growth and damping rates of resonant whistler-mode waves). As shown, the loss-cone driven instability occurs within specific $k_\parallel$-ranges through the $n=-1$ resonance in both cases, overcoming the wave damping through the $n=+1$ and $0$ resonances in the range of the propagation angle to with respect to $\mathbf{B}_0$ from $\theta=0.01^\circ$ to $\theta=30^\circ$. However, at higher $\theta$, the wave damping dominates over the wave instability in the whole $k_\parallel$-space.

Fig.~\ref{energyresult} and Fig.~\ref{energyresulthigh}, corresponding to results from soft and hard populations of relativistic electrons respectively, illustrate Eq.~(\ref{finalenergyntotal}) (i.e., illustrate the energy density of whistler-mode waves associated with the kinetic energy of relativistic electrons through the $n=+1$, $-1$ and $0$ resonances). Under the action of the loss-cone driven instability, the kinetic energy of the relativistic and resonant electrons transfers to the unstable whistler-mode waves through the $n=-1$ resonance. This is the reason why an explicit increase in $W_{\text{Particle}}^{R}$ from both Fig.~\ref{energyresult} and Fig.~\ref{energyresulthigh} occurs through the $n=-1$ resonance, within the unstable $k_\parallel$- and $\theta$-ranges where the loss-cone driven instability occurs. From $\theta=30^\circ$, such explicit increase in $W_{\text{Particle}}^{R}$ nearly disappears as if the loss-cone driven instability does, and $W_{\text{Particle}}^{R}$ becomes almost zero in the $k_\parallel$-space. These demonstrate that Eq.~(\ref{finalenergyntotal}) reflects the wave growth implied by Eq.~(\ref{finalgrowthrate}) in the $k_\parallel$-space, as expected. 

The structures of Eqs.~(\ref{finalenergyntotal}) and (\ref{finalgrowthrate}) are same. But, Eq.~(\ref{finalgrowthrate}) for the wave growth/damping rate is only related to the resonant particles while Eq.~(\ref{finalenergyntotal}) for the wave energy density is related to not only resonant but also non-resonant particles. Therefore, Eq.~(\ref{finalenergyntotal}) considers the total wave-plasma system, and there are some different and still puzzling behaviours in Eq.~(\ref{finalenergyntotal}) compared to Eq.~(\ref{finalgrowthrate}). If the plasma is cold (i.e., if $f_j=\overline n_j^{C} f_j^C$), Eq.~(\ref{finalenergyntotal}) is positive semi-definite as shown in Eq.~(\ref{finalcoldenergy}). However, this is not the case for the hot or relativistic plasma. For example, there are some negative contributions to $W_{\text{Particle}}^{R}$ as shown in Figs.~\ref{energyresult} and \ref{energyresulthigh}, which is counterintuitive and still an open question. Note that, in our numerical calculations, the negative contributions to $W_{\text{Particle}}^{R}$ do not lead to a negative wave energy density (i.e., $W<0$ in Eq.~(\ref{rwenergy})) because $W_{\text{Field}}$ and $W_{\text{Particle}}^{C}$ are all positive, and $W_{\text{Field}}\gg |W_{\text{Particle}}^{R}|$ and $W_{\text{Particle}}^{C}\gg |W_{\text{Particle}}^{R}|$.  However, \citet{osti_4663893} argue that the second term of Eq.~(\ref{waveenergy}), which is the origin of Eq.~(\ref{finalenergyntotal}), can lead to the negative wave energy (i.e., $W<0$) in the presence of a temperature anisotropy. It is argued that the concept of the negative wave energy is important to understand the non-linear instability and wave-wave interactions \citep{https://doi.org/10.1029/JA076i031p07527,cairns_1979,Joarder1997AMO}. However, because it is beyond the scope of the present paper to find a clear insight for the concept of the negative wave energy in wave-particle interactions, further investigations for Eq.~(\ref{finalenergyntotal}) are necessary.

%continuously increases due to the $n=+1$ resonance between parallel-propagating whistler waves and electrons when the value of $k_\parallel$ becomes large, in which $p_{\parallel res}^{+}$ is very small.
%\begin{equation}
%    \frac{\partial p_{\parallel res}^{+}}{\partial \omega_k}\approx\frac{m_{0e}}{k_\parallel}\left(1-\frac{n |\Omega_{0e}|+\omega_k}{|\Omega_{0e}|\cos\theta-\omega_k}\frac{|\Omega_{0e}|\cos\theta}{2\omega_k} \right).
%\end{equation}

Eqs.~(\ref{finalenergyntotal}) and (\ref{finalgrowthrate}) allow us to estimate the wave energy density and growth rate of an arbitrary wave with an arbitrary propagation angle in relativistic plasmas, based on any configuration of the gyrotropic particle MDFs. These analytical results based on linear theory can be extended by cooperating with the quasi-linear theory in relativistic plasmas, which explains the time evolution of relativistic particles through an instability in the momentum space. For example, if an initially unstable MDF is given, we first find properties of an unstable wave by using our analytic expressions presented in this paper. Then, secondly, by applying these unstable wave properties to the quasi-linear diffusion model given by \citet{Jeong_2020}, we temporally evolve the initially unstable MDF toward a stable state. We repeat these two steps until the entropy stabilisation is achieved. This approach will be able to show how relativistic electrons in Fig.~\ref{initiallossconeMDF} diffuse into the loss cone with time through the $n=-1$ resonance, and also show the time evolution of resonant whistler-mode wave properties at the same time. 

Therefore, our presentation for the wave energy density and growth rate in relativistic plasmas helps us better understand wave and plasma observations, and kinetic simulation results from wave instabilities. Moreover, Eq.~(\ref{finalenergyntotal}) provides novel insights and interesting open questions in studying the wave-particle interactions. Besides the loss-cone driven instability, Eq.~(\ref{finalenergyntotal}) can be applied to other instabilities such as electron- or ion-driven instabilities in the solar corona and solar wind to study the change of the resonant wave energy across the wavenumber space \citep{Vocks_2003,2006LRSP....3....1M,10.3389/fspas.2022.951628}.

\section*{Acknowledgements}

S-Y. J. and C. W. are supported by Science and Technology Facilities Council (STFC) grant ST/W000369/1. We appreciate helpful discussion with Daniel Verscharen.

%%%%%%%%%%%%%%%%%%%%%%%%%%%%%%%%%%%%%%%%%%%%%%%%%%
\section*{Data Availability}

The data underlying this paper will be shared on reasonable request
to the corresponding author.



\bibliographystyle{mnras}
\bibliography{mnras_template} % if your bibtex file is called example.bib

%\begin{thebibliography}{99}
%\end{thebibliography}

%%%%%%%%%%%%%%%%%%%%%%%%%%%%%%%%%%%%%%%%%%%%%%%%%%

%%%%%%%%%%%%%%%%% APPENDICES %%%%%%%%%%%%%%%%%%%%%

%%%%%%%%%%%%%%%%%%%%%%%%%%%%%%%%%%%%%%%%%%%%%%%%%%

% Don't change these lines
\bsp	% typesetting comment
\label{lastpage}
\end{document}